# Representation of Dormant and Active Microbial Dynamics for Ecosystem Modeling

Running Title: **Dormant and Active Microbial Dynamics**

Gangsheng Wang[1,2*], Melanie A. Mayes[1,2], Lianhong Gu[1,2], Christopher W. Schadt[1,3]

[1]Climate Change Science Institute, Oak Ridge National Laboratory, Oak Ridge, TN 37831-6301 USA

[2]Environmental Sciences Division, Oak Ridge National Laboratory, Oak Ridge, TN 37831-6301 USA

[3]Biosciences Division, Oak Ridge National Laboratory, Oak Ridge, TN 37831-6038 USA

***Corresponding Author**: **Gangsheng Wang**

Bldg 2040, Room E272, MS-6301

Oak Ridge National Laboratory

Oak Ridge, TN 37831-6301

Tel: (865)574-7615; Fax: (865)574-9501; Email: wangg@ornl.gov

**Abstract**: 199 words; **Main text**: 5300 words; 50 **References**; 2 **Tables**; 4 **Figures**

**Abstract**

Dormancy is an essential strategy for microorganisms to cope with environmental stress. However, global ecosystem models typically ignore microbial dormancy, resulting in major model uncertainties. To facilitate the consideration of dormancy in these large-scale models, we propose a new microbial physiology component that works for a wide range of substrate availabilities. This new model is based on microbial physiological states and is majorly parameterized with the maximum specific growth and maintenance rates of active microbes and the ratio of dormant to active maintenance rates. A major improvement of our model over extant models is that it can explain the low active microbial fractions commonly observed in undisturbed soils. Our new model shows that the exponentially-increasing respiration from substrate-induced respiration experiments can only be used to determine the maximum specific growth rate and initial active microbial biomass, while the respiration data representing both exponentially-increasing and non-exponentially-increasing phases can robustly determine a range of key parameters including the initial total live biomass, initial active fraction, the maximum specific growth and maintenance rates, and the half-saturation constant. Our new model can be incorporated into existing ecosystem models to account for dormancy in microbially-mediated processes and to provide improved estimates of microbial activities.

**INTRODUCTION**

Ecologically-important processes such as soil organic carbon and nutrient cycling largely depend on the active fraction of microbial communities (Blagodatsky et al. 2000). At any given time in a given environment, microorganisms can be in active, dormant, or dead states (Mason et al. 1986). When environmental conditions are unfavorable for growth, e.g., resource limitation, microbes may enter a reversible state of low to zero metabolic activity to alleviate the loss of biomass and metabolic functions (Lennon and Jones 2011, Stolpovsky et al. 2011). The maintenance coefficient (i.e., maintenance cost of C per unit microbial biomass C per unit time) can be two to three orders of magnitude lower in dormant microbes than in metabolically active microbes (Anderson and Domsch 1985a, b). Dormancy is considered an evolutionary strategy designed to maintain the genetic code until conditions improve to allow replication (Price and Sowers 2004). Many soils have slow organic matter turnover rates with seasonal changes in substrate supply, temperature, and moisture. The complexity of soils in space and time may result in uneven distributions of multiple potentially limiting resources, leading to significant rates of dormancy even when some resources are abundant. When spatial and temporal complexity is combined with differential resource partitioning among species in a community, high rates of dormancy could be a prominent feature in soil systems. Thus it is essential to understand dormancy in order to predict the active fraction of microbial communities.

A complicating factor in studying microbial dormancy is that no single approach can simultaneously measure individual microbial states (active, dormant or dead), and a combination of different techniques is required. Differential staining is often used to segregate physiological states with direct microscopic counting of bacteria and fungi. 'Life-indicating' stains that require the presence of 'standard' physiological abilities, such as the esterase activity needed for fluorescein diacetate cleavage, may distinguish active from dormant+dead cells (Adam and Duncan 2001). When combined with general-purpose stains, these strains can distinguish dormant cells by difference (Jones and Senft 1985). Combining membrane-permeant with membrane-impermeant nucleophilic stains (e.g., SYTO-9 and propidium iodide respectively) may distinguish live from dead, but not active from dormant (Boulos et al. 1999, Stocks 2004). Active microbes may or may not be 'viable' with common culture-based techniques, which complicates classification and measurement of dormancy phenomena (Lennon and Jones 2011). Methods such as direct plating, serial dilution and most probable number (MPN) techniques will not distinguish between active and dormant organisms (Schulz et al. 2010). Substrate Induced Respiration (SIR) or Substrate Induced Growth Response (SIGR) methods (Anderson and Domsch 1978, Colores et al. 1996) can distinguish active and dormant communities if growth respiration curves are modeled (using initial exponentially-increasing respiration); however, the technique must be combined with microscopy or chloroform fumigation/extraction in order to obtain total live microbial biomass (Jenkinson and Powlson 1976, Lodge 1993).

Despite limitations in establishing the active biomass, abundant evidence indicates that the majority of environmental microorganisms in a given community may be dormant under natural conditions (Blagodatsky et al. 2000, Yarwood et al. 2013). Alvarez *et al*. (1998) reported that only 3.8–9.7% of the total biomass is active in a Typic Argiudoll soil from the Argentinean Pampa. Khomutova *et al*. (2004) showed that the fraction of active microbial biomass ranged from 0.02% to 19.1% in the subkurgan paleosoils of different age and 9.2–24.2% in modern background soils. Microbial biomass measured through SIR method is thought to reflect only the active portion because the maintenance respiration of dormancy biomass is negligible in the initial exponentially-increasing phase (Lodge 1993, Colores et al. 1996, Orwin et al. 2006).

Through a mathematical analysis of respiration curves, Van de Werf & Verstraete (1987) examined 16 soils and found that 4–49% of the total biomass was in an active state; and the active component in undisturbed natural ecosystems (18.8±8.8%, mean±standard deviation) was about 70% of that in arable agricultural soils (25.7±14.8%). Stenström *et al*. (2001) showed that the fraction of active biomass typically varied from 5% to 20% in soils with no recent addition of substrates. Lennon & Jones (2011) found much lower active fractions in soils (18±15%) than in marine (65±19%) and fresh (54±11%) water environments. If the studies cited above represent a general pattern, then the active fraction is likely below 50% of live microbes under natural soil conditions.

Microbially-mediated processes have been incorporated into ecosystem models (Schimel and Weintraub 2003, Lawrence et al. 2009, Moorhead et al. 2012, Sinsabaugh et al. 2013, Wang et al. 2013) although continued development is still required to bring microbial processes into global climate models (Todd-Brown et al. 2012, Treseder et al. 2012, Wieder et al. 2013). However, these recent models do not consider physiological state changes and assume that measures of microbial biomass constitute the active biomass. Generally, there are two strategies to represent the physiological state in microbial-ecology models: one strategy is to explicitly separate the total live biomass into two pools, i.e., active and dormant (e.g., Konopka 1999, Stolpovsky et al. 2011); the other is to directly regard the active fraction (i.e., ratio of active biomass to total live biomass) as a state variable (e.g., Panikov 1996, Blagodatsky and Richter 1998). These two approaches are equivalent since they both predict the total live biomass, active and dormant biomass, and the flux or net flux between the active and dormant components. The above-mentioned modeling efforts have shown that adequate representation of dormancy and the transition between the dormant and active fractions is crucial for modeling important microbially-mediated ecosystem processes.

Here, we review state-of-the-art microbial dormancy modeling approaches and discuss the rationales of these models with a focus on transformation processes between active and dormant states. We propose an improved synthetic microbial physiology model based on accepted assumptions and examine the model behavior with theoretical and experimental analyses. In this paper, the 'total microbial biomass' refers to the 'total live microbial biomass' unless otherwise stated. Our objective is to clarify the applicability of existing microbial dormancy models and provide a new theoretical basis for representing microbial activity and dormancy in ecosystem models.

## DORMANCY IN MICROBIAL MODELS

### Transformation between active and dormant states

Although Buerger *et al*. (2012) argued that dormant microbial cells could reactivate stochastically and might be independent of environmental cues, environmental factors such as substrate availability are often thought to control the transformation between active and dormant states (Lennon and Jones 2011). Most models (see Appendix S1 in Supporting Information for a summary) distinguish the active biomass pool from the dormant pool and define them as two state variables ($B_a$ and $B_d$) (see Fig. 1). Only active microbes ($B_a$) can uptake substrate and reproduce new cells. The connection between the active and dormant states is a reversible process including two directional sub-processes, i.e., dormancy (from active to dormant) and reactivation (or resuscitation, from dormant to active). Losses from active biomass include growth respiration and maintenance (maintenance respiration, mortality, enzyme synthesis, etc.)

(Wang et al. 2013). Dormant microbes still require energy for maintenance and survival although at a lower metabolic rate (Lennon and Jones 2011).

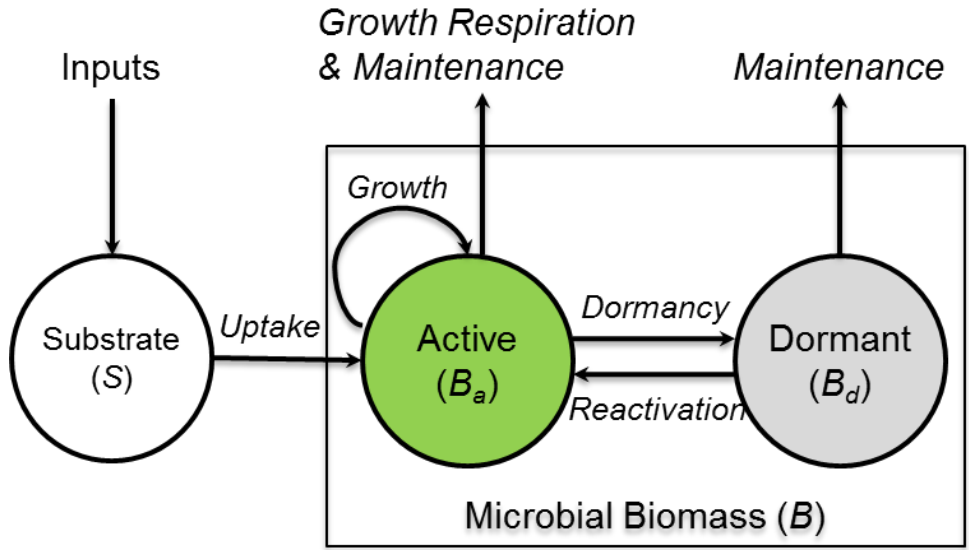

**Figure 1.** Active and dormant biomass pools in microbial physiology models (modified from Fig. 2 in Lennon & Jones, 2011)

The net transformation rate ($B_{a \to d}^{N}$) from active to dormant state is the difference between the flux from active to dormant ($B_{a \to d}$) and the flux from dormant to active state ($B_{d \to a}$), i.e., $B_{a \to d}^{N} = B_{a \to d} - B_{d \to a}$. The models of Hunt (1977) and Gignoux *et al.* (2001) directly formulate the net flux ($B_{a \to d}^{N}$) without explicit components for $B_{a \to d}$ and $B_{d \to a}$. The direction of the net flux depends on the maintenance requirement relative to the substrate availability. If the available substrate is less than the maintenance requirement, there is a positive net flux from active to dormant pool, and vice versa. In addition, Hunt (1977) assumed a "buffer zone" for the change of states: when the maintenance requirement surpasses the substrate supply but the deficit is within a small fraction (1% $d^{-1}$) of $B_a$, there is no flux between the two states.

Some models define rates for both dormancy and reactivation. In the model of Ayati (2012), the dormant rate ($\gamma_{a \rightarrow d}$) increases with declining substrate concentration, and the reactivation ($\gamma_{d \rightarrow a}$) only occurs when substrate concentration is higher than the half-saturation constant ($K_s$). Konopka (1999) modified the potential rates for deactivation and reactivation by the relative growth rate ($\mu/\mu_{max}$, ratio of true specific growth rate to maximum specific growth rate), i.e, the two rates are multiplied by ($1-\mu/\mu_{max}$) and $\mu/\mu_{max}$, respectively. Similarly, Jones & Lennon (2010) postulated two complementary rates ($1-R$ and $R$) for dormancy and resuscitation.

Two other models also explicitly formulate the two conversion rates between states but do so using concepts of "probability". Bär *et al*. (2002) used two complementary factors ($1-J$ and $J$) to represent the probability for the transition between active and dormant state in addition to an identical potential rate constant for the two processes. The conceptual model of Locey (2010) applies a deterministic dormant rate and a stochastic resuscitation rate. The potential resuscitation rate is modified by ($1-p$), where $p$ is the probability that a disturbance in the active pool will result in the immigration of one individual from the metacommunity. The probability ($J$) in Bär *et al*. (2002) is explicitly calculated from the environmental cues (e.g., soil moisture), while the cause of the probability ($p$) in Locey (2010) is not elucidated.

**Switch function model**

In addition to the dormancy and reactivation processes, a key concept in the model (see Appendix S1) developed by Stolpovsky *et al*. (2011) is 'switch function ($\theta$)'. The switch function determines the fraction of active cells taking up dissolved organic carbon (DOC). This function refers to the growth fraction in active biomass ($B_a$) that consumes substrate and thus is not the same as the active fraction ($r$) in total biomass ($B$). Furthermore, the dormancy and reactivation fluxes are set to be proportional to (1−θ) and θ, respectively. θ is formulated by the Fermi-Dirac statistics (Stolpovsky et al. 2011). Another feature of this model is the consideration of "depth" of dormancy in reactivation, where the reactivation rate is negatively dependent on the duration of dormancy. The switch function model has a detailed description of DOC consumption and microbial processes. However, with at least 15 model parameters, the application of this model may suffer from 'over-parameterization' (Reichert and Omlin 1997). In addition, it is difficult to compute the Gibbs energy change of the oxidation of DOC (Stolpovsky et al. 2011).

We believe the inclusion of the switch function in DOC consumption and microbial growth results in double counting of the impact of substrate and terminal electron acceptor (TEA). According to the Michaelis-Menten (M-M) kinetics, the substrate saturation level represents the fraction of enzyme-substrate complex ($ES$) in active enzyme ($E_0$), where the substrate saturation level is formulated by $S/(K_s + S)$ with $S$ and $K_s$ being the substrate concentration and the half-saturation constant (Wang and Post 2013). When the M-M (or Monod) kinetics is applied to describe microbial uptake of substrate, the substrate (or combined with TEA) saturation level is a measure of the actively growing fraction in the active microbial community. The switch function is also determined by the saturation levels of substrate and TEA, i.e., $\mu(S, TEA)$ (see Appendix S1). Therefore, either $\theta$ or $\mu(S, TEA)$ may be used to modify the microbial uptake rate but the inclusion of both is not only unnecessary but also inappropriate.

**Physiological state index models**

As an alternative to models with two microbial biomass pools (i.e., active and dormant), a further state variable indicating the dormant or active fraction in total biomass has been proposed.

Wirtz (2003) developed a simple index ($r_d$ = 0.5–1.0) representing the dormant microbial biomass as a fraction of the steady-state total biomass ($B_{stat}$) under the condition of $B_d \ll B_a$. In case of a net loss of total biomass ($dB/dt < 0$), the dormant biomass $B_d = B_{stat} \cdot r_d$; otherwise ($dB/dt > 0$), $B_d = B_{stat} \cdot (1 - r_d)$. This model has a sudden change of dormant biomass at the transition point (i.e., $dB/dt = 0$) since $r_d > 0.5$.

Different from the dormant index of Wirtz (2003), the concept of an active index (i.e., index of physiological state) of soil microbial community has been employed in soil carbon and nutrient cycling models (Panikov 1996, Blagodatsky and Richter 1998). The index of physiological state ($r$), referring to the activity state, is often defined as the ratio of metabolically active microbial biomass to the total soil microbial biomass (Panikov 1996, Blagodatsky and Richter 1998, Stenström et al. 2001).

In the Synthetic Chemostat Model (SCM), the rate of change of the state variable $r$ is described as follows (Panikov 1995, 1996):

$$\frac{dr}{dt} = \frac{1}{B}\frac{dB}{dt} \cdot (\phi - r) = \mu \cdot (\phi - r) \qquad (1)$$

with $\phi = \phi(S) = S^n/(K_r + S^n)$, or $\phi = S/(K_r + S)$ (2)

where $r = B_a/B$, representing the fraction (hereinafter referred to as 'active fraction') of active biomass in total biomass; $\mu$ is the specific growth rate of total biomass; $\phi$ denotes the saturation level of substrate ($S$); the simple power ($n = 1$) has been widely used (Panikov and Sizova 1996) and, in this case ($n = 1$), $K_r$ is called the half-saturation constant.

Blagodatsky & Richter (1998) used the expression $\mu(S) = \mu_{max} \cdot \phi(S)$ in their model development. This expression was not derived in the original definition of the specific growth rate by Panikov (1995) and because its validity cannot be inferred, the concepts will not be addressed here.

According to Panikov's derivation (Panikov 1995), the specific growth rate ($\mu$) follows the general definition (Pirt 1965, Wang and Post 2012):

$$\mu = \frac{1}{B}\frac{dB}{dt} \qquad (3)$$

Based on Eqs. 1 and 3, we can derive (see Appendix S2):

$$dB_a/dt = \phi \cdot (dB/dt) \qquad (4)$$

$$dB_d/dt = (1-\phi) \cdot (dB/dt) \qquad (5)$$

We find that the model described by Eq. 1 is not applicable under low substrate conditions, as described below. Generally, the rates of change in biomass pools ($B$, $B_a$, and $B_d$) can be expressed as

$$dB/dt = g^{\pm}(S, B_a) - f^{+}(S, B_d) \qquad (6)$$

$$dB_a/dt = g^{\pm}(S, B_a) - B_{a \to d}^{N} \qquad (7)$$

$$dB_d/dt = -f^{+}(S, B_d) + B_{a \to d}^{N} \qquad (8)$$

where $B_{a \to d}^{N}$ denotes the net dormancy flux; $g^{\pm}(S, B_a)$ is a function that represents the growth and maintenance of $B_a$, i.e., the net growth of $B_a$; and $f^{+}(S, B_a)$ is a function denoting the maintenance and survival energy costs of $B_d$. The superscript '$\pm$' in $g^{\pm}$ indicates the function value of $g$ may be positive (at high $S$) or non-positive (at low $S$) since both growth and maintenance occur in $B_a$. The superscript '+' in $f^{+}$ implies $f \geq 0$. Note that the function $f^{+}(S, B_a)$ is not necessarily dependent on $S$.

From Eqs. 4, 6 and 7, we can obtain

$$B_{a \to d}^{N} = (1-\phi) \cdot g^{\pm}(S, B_a) - \phi \cdot f^{+}(S, B_d) \qquad (9)$$

The two terms in the right side of Eq. 9 may be regarded as the conversion of $B_a$ to $B_d$ (i.e., $B_{a \to d}$) and the transformation of $B_d$ to $B_a$ (i.e., $B_{d \to a}$), respectively. At high $S$ resulting in $g \geq 0$, Eq. 9 may be one of the possible expressions for $B_{a \to d}$ and $B_{d \to a}$. However, at low $S$ leading to $g < 0$ and $B_{a \to d} < 0$, i.e., no active cells become dormant under insufficient substrate, which is inconsistent with the strategy of dormancy for microorganisms when faced with unfavorable environmental conditions (Lennon and Jones 2011).

Based on the above analysis, we conclude that the physiological state index model (Eq. 1) can be improved. In other words, the empirical assumption that the steady state active fraction ($r^{ss}$) approaches the substrate saturation level ($\phi^{ss}$) may not be necessary because this assumption could lead to impractical flux (Eq. 9) between dormant and active states under low substrate conditions.

## A SYNTHETIC MICROBIAL PHYSIOLOGY MODEL

Based on the aforementioned review and analysis, we have developed a synthetic microbial physiology model component relating to substrate availability.

### General assumptions

First we define the substrate saturation level ($\phi$) as

$$\phi = S/(K_S + S) \qquad (10)$$

where the parameter $K_s$ is the half saturation constant for substrate uptake as indicated by the Michaelis-Menten kinetics (Wang and Post 2013).

Based on the above review of existing dormancy models, the following assumptions are accepted in our new model: (1) the dormancy rate is proportional to the active biomass and the reactivation rate is proportional to the dormant biomass, i.e., $B_{a \to d} \propto B_a$ and $B_{d \to a} \propto B_d$; (2)

under very high substrate concentration ($S \gg K_s$), $\phi \to 1$, $B_{a \to d} \to 0$ and $B_{d \to a} \geq 0$; (3) under very low substrate ($S \ll K_s$), $\phi \to 0$, $B_{a \to d} \geq 0$ and $B_{d \to a} \to 0$; (4) based on the assumptions (1–3), we derive that $B_{a \to d} \propto (1-\phi) \cdot B_a$ and $B_{d \to a} \propto \phi \cdot B_a$; (5) further we assume that the maximum specific maintenance rate ($m_R$ with units of h$^{-1}$) controls both transformation processes since the maintenance energy cost is the key factor regulating the dormancy strategy (Hunt 1977, Gignoux et al. 2001, Lennon and Jones 2011). Thus we postulate that

$$B_{a \to d} = (1-\phi) \cdot m_R \cdot B_a \qquad (11a)$$

$$B_{d \to a} = \phi \cdot m_R \cdot B_d \qquad (11b)$$

**Model description**

Combining Eqs. 11a and 11b with the MEND model (Wang and Post 2012, Wang et al. 2013), we express the microbial physiology component (see Fig. 1) as a group of differential equations

$$dS/dt = I_s - \frac{1}{Y_G} \cdot \frac{\phi}{\alpha} m_R \cdot B_a \qquad (12a)$$

$$dB/dt = d(B_a + B_d)/dt = (\phi/\alpha - 1) \cdot m_R \cdot B_a - (\beta \cdot m_R) \cdot B_d \qquad (12b)$$

$$dB_a/dt = (\phi/\alpha - 1) \cdot m_R \cdot B_a - (1-\phi) \cdot m_R \cdot B_a + \phi \cdot m_R \cdot B_d \qquad (12c)$$

$$dB_d/dt = -(\beta \cdot m_R) \cdot B_d + (1-\phi) \cdot m_R \cdot B_a - \phi \cdot m_R \cdot B_d \qquad (12d)$$

where $t$ is the time scale; $\phi$ is defined by Eq. 10; $I_s$ is the input to substrate pool; $Y_G$ is the true growth yield; $m_R$ denotes the specific maintenance rate at active state (h$^{-1}$); $\alpha = m_R/(\mu_G + m_R)$ is the ratio of $m_R$ to the sum of maximum specific growth rate ($\mu_G$) and $m_R$, $\alpha \in (0, 0.5)$ since usually $m_R < \mu_G$; and $\beta$ (0–1) is the ratio of dormant maintenance rate to active maintenance rate, i.e., ($\beta \cdot m_R$) denotes the maximum specific maintenance rate at dormant state.

In summary, there are five parameters ($\alpha$, $\beta$, $m_R$ or $\mu_G$, $Y_G$, $K_s$) in the proposed model (hereinafter referred to as MEND model). From Eqs. 12b and 12c, we can derive the change rate of active fraction ($r$) (see Appendix S2)

$$dr/dt = m_R \cdot \left[(\phi - r) + (\phi/\alpha + \beta - 1) \cdot r \cdot (1 - r)\right] \quad (12e)$$

This equation for $r$ is more complicated than Eq. 1 but still practical, given currently available data. Additionally, it implies that $r$ is not necessary to approach $\phi$ at steady state.

**Steady state analysis**

Assuming the input ($I_s$) is time-invariant, we can obtain the steady state solution to the MEND model (see Appendix S2). Fig. 2 shows the steady state active fraction ($r^{ss}$) and substrate saturation level ($\phi^{ss}$) as a function of the two physiological indices, i.e., $\alpha$ (0–0.5) and $\beta$ (0–1). Both $r^{ss}$ and $\phi^{ss}$ positively depend on $\alpha$ and $\beta$ and $r^{ss} \geq \phi^{ss}$ for any combinations of $\alpha$ and $\beta$. If we consider two extreme values of $\beta \to 0$ or $\beta \to 1$, the $r^{ss}$ and $\phi^{ss}$ (see Appendix S2) can be simplified to

$$r^{ss}_{\beta \to 0} = \phi^{ss}_{\beta \to 0} = \alpha \quad (13a)$$

$$\begin{cases} r^{ss}_{\beta \to 1} = \dfrac{1 + \sqrt{1 + 8\alpha}}{4} \\ \phi^{ss}_{\beta \to 1} = \dfrac{4\alpha - 1 + \sqrt{1 + 8\alpha}}{3 + \sqrt{1 + 8\alpha}} \end{cases} \quad (13b)$$

Eq. 13 and Fig. 2 indicate that: (1) the steady state active fraction ($r^{ss}$) is equal to $\phi^{ss}$ and they are identical to $\alpha = m_R/(\mu_G + m_R)$ only under the condition of $\beta \to 0$; (2) the upper bound of $r^{ss}$ is approximately 0.8 at $\alpha \to 0.5$ and $\beta \to 1$; and (3) with $\alpha \leq 0.5$, the maximum $r^{ss}$ is at the level of 0.5 if the magnitude of $\beta$ is around 0.001–0.01 (Anderson and Domsch 1985a). This threshold value (0.5) of $r^{ss}$ is a reasonable estimate that can explain how the measured active fraction of microbes in undisturbed soils is usually considerably less than the total biomass (Van de Werf and Verstraete 1987, Stenström et al. 2001, Lennon and Jones 2011).

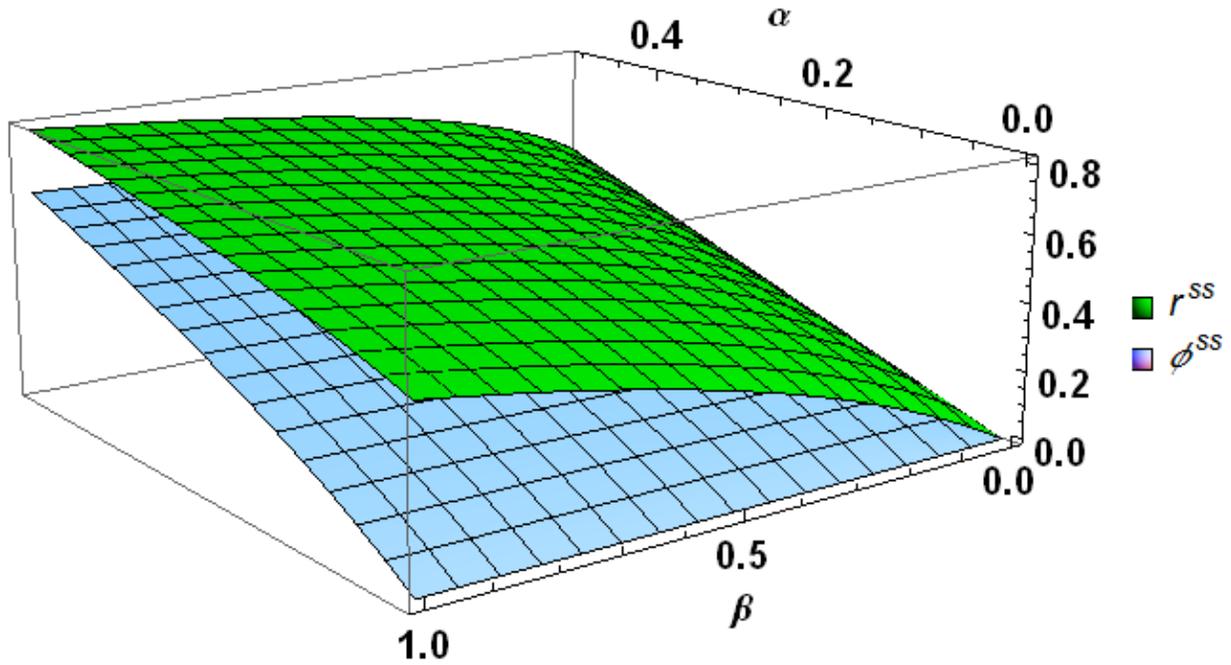

**Figure 2.** Steady state active fraction ($r^{ss}$) and substrate saturation level ($\phi^{ss}$) as a function of $\alpha$ and $\beta$; $\alpha = m_R/(\mu_G + m_R)$, $\mu_G$ and $m_R$ ($h^{-1}$) are maximum specific growth rate and specific maintenance rate for active biomass, respectivly; $\beta$ denotes the ratio of dormant specific maintenance rate to $m_R$.

**Model simplification under sufficient substrate condition**

The simplification of the microbial model under excess substrate has been employed to estimate maximum specific growth rate ($\mu_G$), active microbial biomass ($B_a$), and/or total microbial biomass ($B$) using the SIR or SIGR data (Colores et al. 1996, Panikov and Sizova 1996,

Blagodatsky et al. 2000). Here we also show the simplification of our model (Eq. 12) for conditions appropriate to SIGR or SIR experiments, e.g., the short-term period of exponentially-increasing respiration of active biomass following substrate addition. We will test our reduced and full model with the SIGR data of Colores *et al.* (1996) in the next section.

Under sufficient substrate (i.e., $S \gg K_s$ in Eq. 10 thus $\phi \rightarrow 1$), Eqs. 12(a–e) can be simplified and integrated for initial conditions, i.e., $S = S_0$, $B = B_0$ and $r=r_0$ at $t = 0$ (see Appendix S2):

$$S(t) = S_0 - \frac{B(t) - B_0}{Y_G(1-\alpha)} \tag{14a}$$

$$B(t) = B_0 r_0 \cdot e^{\mu_G t} + B_0(1-r_0) \cdot [\alpha \cdot e^{\mu_G t} + (1-\alpha) \cdot e^{-m_R t}] \tag{14b}$$

$$r(t) = \frac{[r_0 + \alpha(1-r_0)] \cdot e^{(\mu_G + m_R)t} - \alpha(1-r_0)}{[r_0 + \alpha(1-r_0)] \cdot e^{(\mu_G + m_R)t} + (1-\alpha)(1-r_0)} \tag{14c}$$

The $CO_2$ production rate, $v(t)$, during the exponential growth stage is derived as an explicit function of $t$ (see Appendix S2):

$$v(t) = \frac{dCO_2}{dt} = \frac{B_0(1-Y_G)}{Y_G} \left\{ [(\mu_G + m_R) \cdot r_0 + m_R \cdot (1-r_0)] \cdot e^{\mu_G t} - [m_R \cdot (1-r_0)] \cdot e^{-m_R t} \right\} \tag{14d}$$

The respiration rate, $v(t)$, is associated with two exponential items, i.e., $e^{\mu_G t}$ and $e^{-m_R t}$. Considering an extreme case that $m_R \ll \mu_G$ (i.e., $\alpha \rightarrow 0$), Eqs. 14b–14d can be further simplified to Eqs. S2-8b–8d (see Appendix S2).

Eqs. S2-8b and S2-8c (denoting $B(t)$ and $r(t)$, respectively) are similar to Eqs. 11 and 10 in Panikov & Sizova (1996), respectively. However, Eq. S2-8d (denoting $v(t)$) is different from Eq. 13 of Panikov & Sizova (1996), where a constant '*A*' was added to the exponential. Eq. S2-8d is also identical to Eq. 7 derived for SIGR experiments in Colores *et al.* (1996).

Panikov & Sizova (1996) used their Eq. 13 to fit respiration rates during the exponentially-increasing (i.e., no substrate limitation) phase (see Fig. 2 in Panikov & Sizova (1996) for data and curve fittings). However, these data are based on glucose-induced respiration that includes both basal respiration of native SOC and respiration due to the addition of glucose (Colores et al. 1996). The basal respiration rate may be regarded as a constant in certain cases (see Colores *et al*. (1996) and data in Fig.1 of Blagodatsky *et al*. (1998)). The constant '*A*' representing the basal respiration rate was included in Eq. 13 of Panikov & Sizova (1996) in order to fit the combined respiration from the addition of glucose and basal respiration. However, this constant '*A*' cannot be derived from such governing equations as Eqs. S2-6a–6c (see Appendix S2) that assume respiration is the sole result of substrate addition. In other words, the equations do not include basal respiration. Certainly, the predicted respiration could include basal respiration as long as (i) a basal respiration rate is added to Eq. 14d *ad hoc* or (ii) Eqs. S2-6a–6c (or, more commonly, Eqs. 12a–12e) are linked to a soil organic matter (SOM) decomposition model, which can produce decomposed native soil C in addition to the respiration of the substrate addition. Because Eq. 13 of Panikov & Sizova (1996) is not linked to a native C decomposition model, fitting the model to combined native C and substrate respiration data is not appropriate.

**Model test I: substrate-induced respiration**

In this section, we used the respiration data from $^{14}$C-labeled glucose SIGR experiments by Colores et al. (1996) to validate our MEND model. The respiration data only represented the $CO_2$ production from the added substrate and did not include basal respiration from the native C.

First we employed Eq. 14d to fit the respiration rates during the exponentially-increasing stage and the result is shown in Fig. 3a (see original data in Fig. 3 of Colores et al. (1996)). The true growth yield ($Y_G$) was set to 0.5 according to Colores *et al*. (1996). There are four undetermined parameters ($B_0$, $r_0$, $\mu_G$, $\alpha$) in Eq. 14d (with $m_R = \mu_G \cdot \alpha /(1-\alpha)$). We found that only the maximum specific growth rate ($\mu_G$) could be determined with high confidence (coefficient of variation (CV) = 5%) from the exponentially-increasing respiration rates. The CVs of the other three optimized parameters ($B_0$, $r_0$, $\alpha$) were as high as 55–77% (Table 1). However, the initial active microbial biomass ($B_{a0} = B_0 \times r_0$) had a lower uncertainty (CV = 20%) compared to $B_0$ and $r_0$. The above results indicate that the exponentially-increasing respiration rates can only be used to obtain $\mu_G$ and $B_{a0}$.

We then conducted numerical simulations in terms of all data including both exponentially-increasing and non-exponentially-increasing respiration rates (Fig. 3b). The non-exponentially-increasing respiration rates include the lag period before the exponentially-increasing phase and the respiration at longer times after the rates cease to increase exponentially (Colores et al. 1996). The latter phase is likely because of the substrate saturation levels ($\phi$) become limiting to respiration. We used Eqs. 12a, 12b, 12e and the corresponding expression for $CO_2$ flux rate, to allow the substrate saturation level ($\phi$) to change with time. Additionally, we used the ranges of $\mu_G$ determined above. We used the SCEUA (Shuffled Complex Evolution at University of Arizona) algorithm (Duan et al. 1992, Wang et al. 2009) to determine model parameters.

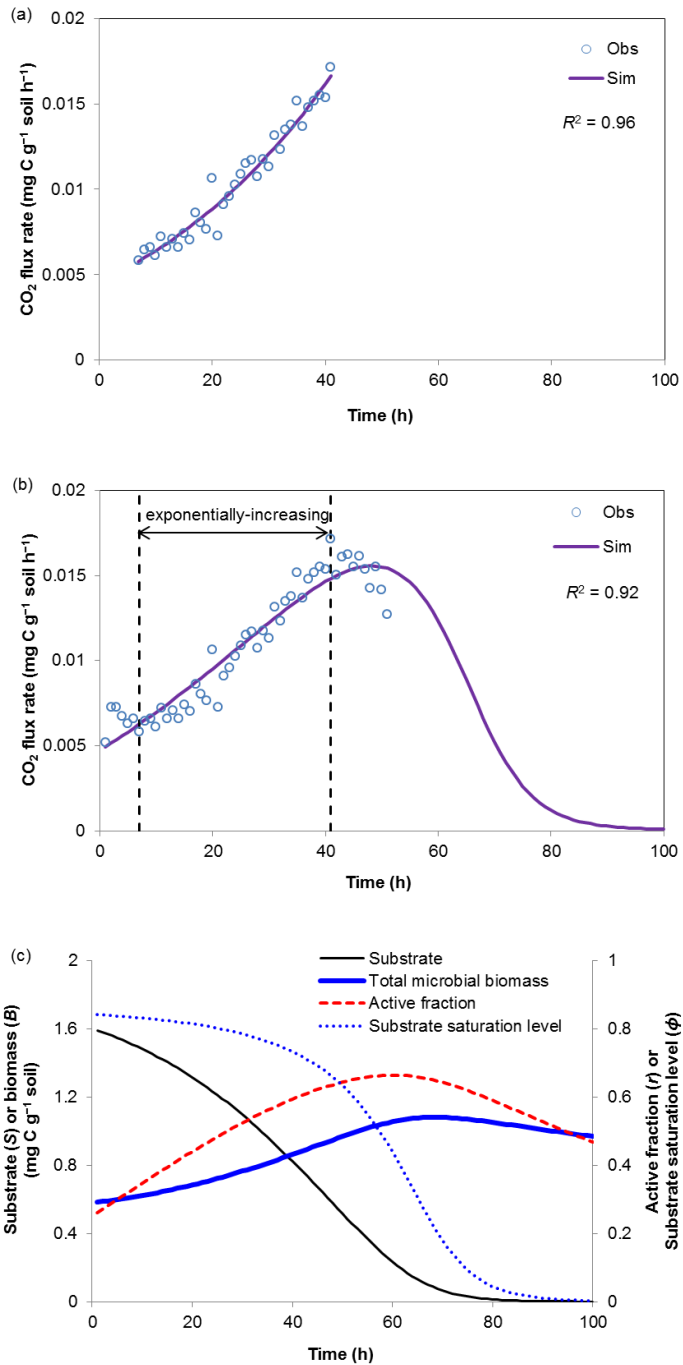

**Figure 3.** MEND model simulations against the respiration rates due to added [14]C-labeled glucose in Colores et al. (1996). (a) Fitting of the respiration rates in the exponentially-increasing phase using eqn 14, 'Obs' and 'Sim' denote observed and simulated data, respectively. (b) Fitting of the respiration rates in both exponentially-increasing and non-exponentially-increasing phases using eqn 12. (c) Simulated substrate ($S$), total live microbial biomass ($B$), active fraction ($r$) and substrate saturation level ($\phi$) based on eqn 12.

**Table 1** MEND model parameters values used for simulation of respiration rates due to added $^{14}$C-labeled glucose in Colores et al. (1996)

| Parameter | Exponentially-increasing respiration* | | | All data† | | | Description |
|---|---|---|---|---|---|---|---|
| | Mean | SD‡ | CV§ | Mean | SD | CV | |
| $B_0$ | 0.504 | 0.279 | 55% | 0.525 | 0.080 | 15% | Initial microbial biomass, (mg C g$^{-1}$ soil) |
| $r_0$ | 0.394 | 0.263 | 67% | 0.285 | 0.064 | 23% | Initial active fraction |
| $\mu_G$ | 0.027 | 0.001 | 5% | 0.030 | 0.001 | 3% | Maximum specific growth rate ( h$^{-1}$) |
| $\alpha$ | 0.185 | 0.142 | 77% | 0.228 | 0.031 | 13% | $m_R/(\mu_G+ m_R)$, $m_R$ is maximum specific maintenance rate for active microbes ( h$^{-1}$) |
| $K_s$ | — | — | — | 0.275 | 0.038 | 14% | Half-saturation constant for substrate (mg C g$^{-1}$ soil) |
| $\beta$ | — | — | — | 0.025 | 0.019 | 76% | Ratio of dormant maintenance rate to $m_R$ |
| $Y_G$ | 0.5 | — | — | 0.5 | — | — | True growth yield, constant |
| $B_{a0}$ | 0.135 | 0.027 | 20% | 0.145 | 0.004 | 3% | Initial active biomass (mg C g$^{-1}$ soil), calculated by $B_0 \times r_0$ |

*Only the respiration rates during exponentially-increasing phase are used.

†All data including both exponentially-increasing and non-exponentially-increasing respiration.

‡SD: standard deviation.

§CV: Coefficient of variation.

When exponentially-increasing and non-exponentially-increasing data are included together, the CVs of all parameters ($B_0$, $r_0$, $\mu_G$, $\alpha$, $K_s$, $\beta$) are within 25% except $\beta$ with a high CV of 76% (Table 1). The optimized $\mu_G$ values (0.030±0.001 h$^{-1}$) are almost the same as obtained by the SIGR method (Colores et al. 1996). Model estimates of $\alpha$ = 0.228±0.031 indicate that the maximum specific maintenance rate of active microbes ($m_R$) is about 30% of $\mu_G$ and thus cannot be ignored. The initial active biomass ($B_{a0}$) is 0.145±0.004 mg C g$^{-1}$ soil (see Table 1), which is lower than the values (0.194±0.004 mg C g$^{-1}$ soil) using the SIGR method. This is likely due to the inclusion of maintenance respiration (characterized by $m_R$, see Eq. 14d) in our model even for the exponentially-increasing stage; thus a lower $B_{a0}$ could produce similar $CO_2$ flux to the case with higher $B_{a0}$ that does not include the contributions from maintenance respiration. Our results also show that the initial active fraction ($r_0$) is 28.5±6.4% and $\beta$ is 0.025±0.019. The magnitude of $\beta$ is comparable to the estimation by Anderson & Domsch (1985a, b). In addition, the half-saturation constant ($K_s$) was estimated as 0.275±0.038 mg C g$^{-1}$ soil, which is very close to the values derived from 16 soils by Van de Werf & Verstraete (1987). This $K_s$ value indicates the substrate saturation level ($\phi$) is higher than 0.7 before the transition from exponentially-increasing to non-exponentially-increasing phase (see Fig. 3c). The changes of substrate ($S$), total microbial biomass ($B$) and active fraction ($r$) with time are also shown in Fig. 3c. In conclusion, the five parameters ($B_0$, $r_0$, $\mu_G$, $\alpha$, $K_s$) can be effectively determined using both exponentially-increasing and non-exponentially-increasing respiration rates, whereas $\beta$ may also be determined but with a relatively high uncertainty (CV = 76%) than the other parameters.

Through this experimental analysis, we identified the need for isotopic data to discriminate between basal and substrate-induced respiration. We also discovered that the exponentially-increasing period due to substrate addition can be used to identify only a select set of model parameters (i.e., $\mu_G$ and $B_{a0}$) as also demonstrated by the method of Colores *et al*. (1996). These parameters, however, can be further applied to longer-term respiration experiments to enable fitting to obtain the remainder of model parameters by using our MEND model. Thus, we have found a new and unique solution to identify different parameters as a function of time, and to effectively use isotopic labeling to yield a specific set of model parameters.

**Model test II: intermittent substrate supply**

In order to further validate this additional physiological component in the MEND model, we also tested it against a laboratory experimental dataset with intermittent substrate supply (Stolpovsky et al. 2011). In addition to the substrate, another limiting factor (i.e., oxygen, $O_2$) was included in this study. For this reason, we also introduced one more parameter ($K_o$: half saturation constant for $O_2$) to represent the limitation of $O_2$ on the microbial processes sketched in Fig. 1. Similar to substrates, the saturation level of $O_2$ is computed as $O_2/(O_2+K_o)$, where $O_2$ denotes the concentration of oxygen. The simulated oxygen concentrations by Stolpovsky et al. (2011) were used as an input to our model. We used the SCEUA algorithm to determine the six model parameters in addition to the initial value for active fraction ($r_0$).

A summary of the seven parameters (one of them is $r_0$) and their fitted values is presented in Table 2. The initial active fraction ($r_0$) has a median of 0.925 with the 95% confidence interval (CI) of [0.628−1.000]. It means that a high $r_0$ is required for this experiment, but not necessary to be 1.0 set by Stolpovsky et al. (2011). The model and data are not sensitive to $β$ since its 95% CI covers a wide range from 0.001 to 1. The reason is that the experiment only lasts for a very short time (33 h) so the influence of low metabolic rate at dormant state is insignificant.

Fig. 4 shows that the simulated total biomass ($B$) and substrate ($S$) concentrations agree very well with the observations (the coefficients of determination are 0.98 and 0.78 for biomass in Fig. 4a and substrate in Fig. 4b, respectively). Our simulation results indicate that, under limited $O_2$ between 12h and 24h of the experiment, the active biomass decreases and the dormant biomass increases. As a result, the active fraction ($r$) declines from ca. 0.9 to 0.7 (Fig. 4a). For the same period Stolpovsky et al. (2011) predicted a decrease of $r$ from 1.0 to ca. 0, which means that all active biomass becomes dormant. Although there were not adequate measurements to confirm either prediction, our predicted changes in the active fraction (*r*) appear to be more reasonable during such a short experimental time period. This demonstration also shows that our model is capable of producing reasonable change in total, active, and dormant microbial biomass in response to substrate supply as well as an important forcing function ($O_2$).

**Table 2** MEND model parameter values used for simulation of the experiment described in Fig. 3 of Stolpovsky et al. (2011)

| Parameter | Fitted Value* | Initial Range | Description |
|---|---|---|---|
| $m_R$ | 0.032, [0.011–0.048] | 0.001–0.1 | Specific maintenance rate for active biomass ($h^{-1}$) |
| $\alpha$ | 0.099, [0.045–0.181] | 0.001–0.50 | $m_R/(\mu_G + m_R)$, $\mu_G$ is specific growth rate ($h^{-1}$) |
| $K_s$ | 3.110, [1.387–5.652] | 0.1–9.0 | Half-saturation constant for substrate (mg $L^{-1}$) |
| $Y_G$ | 0.573, [0.463–0.600] | 0.2–0.6 | Growth yield factor (–) |
| $K_o$ | 0.0008, [0.0007–0.001] | 0.005–0.1 | Half-saturation constant for dissolved oxygen (mM) |
| $\beta$ | 0.351, [0.001–1.000] | 0.001–1 | Ratio of dormant maintenance rate to $m_R$ |
| $r_0$ | 0.925, [0.628–1.000] | 0–1 | Initial fraction of active biomass (–) |

*Medians and 95% confidence intervals of the fitted values from 100 optimization runs, i.e., 100 different random seeds are used for the stochastic optimization algorithm.

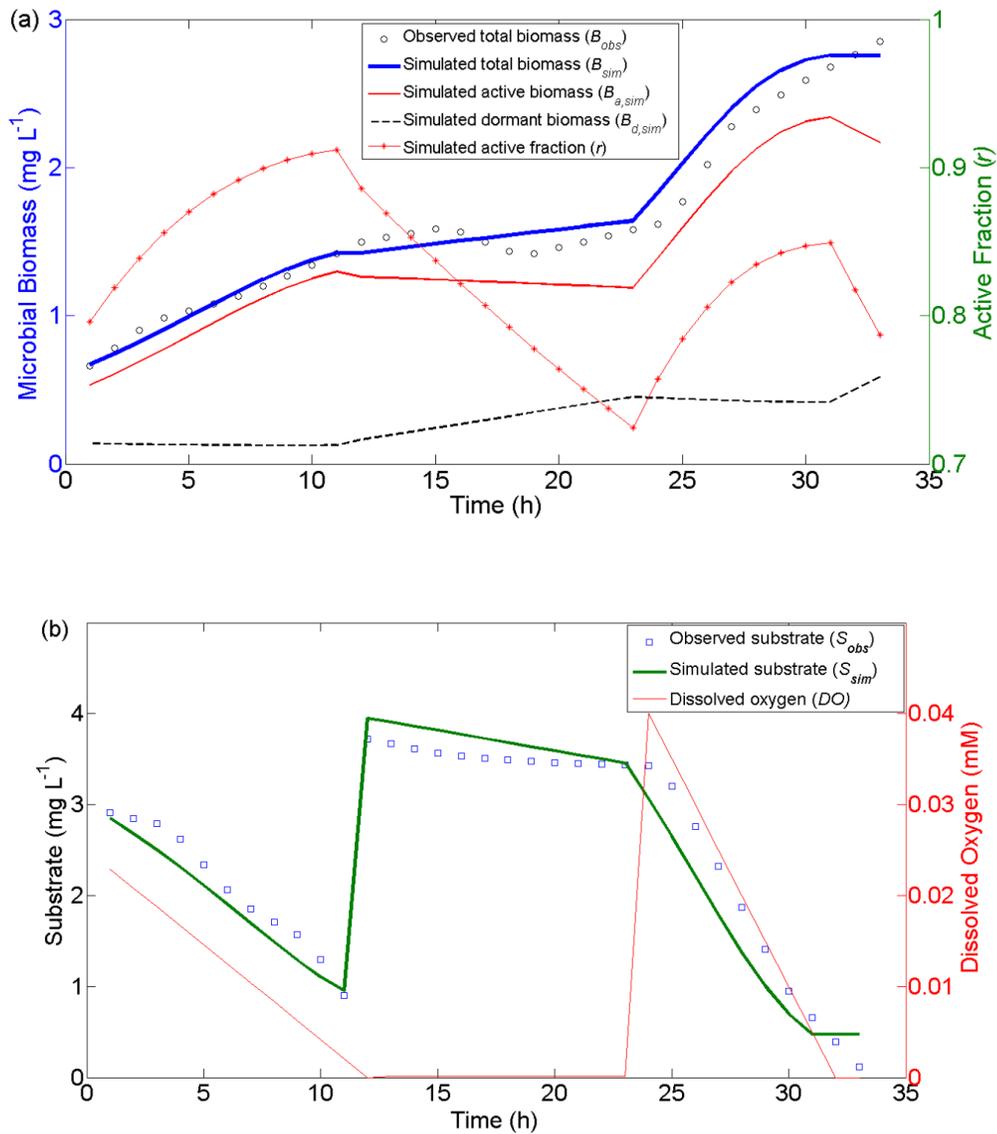

**Figure 4.** MEND model simulations against the experimental dataset used by Stolpovsky et al. (2011). (a) total live biomass, active and dormant biomass, and active fraction; (b) observed and simulated substrate concentration and prescribed $O_2$ concentration. There are three manipulations on the substrate and oxygen: (1) at time 0, the substrate (3 mg/L) and $O_2$ (0.025 mM) are added to the system; (2) after 12 h, the same amount of substrate is injected; (3) at 24 h, additional $O_2$ (0.04 mM) is injected to the system. The observed concentrations of substrate and total biomass are hourly data interpolated from the original observations in Stolpovsky et al. (2011). We scaled the substrate concentrations (with units of mM in original data) to match the magnitude of biomass concentration in units of mg/L.

**CONCLUSION**

We show that the physiological state index model (Eq. 1) of Panikov (1996) can be improved by eliminating the assumption that the steady state active fraction ($r^{ss}$) approaches the substrate saturation level ($\phi^{ss}$). In particular, the model of Panikov (1996) indicates that no active cells become dormant under insufficient substrate, which disregards the general nature of the strategy of dormancy in microorganisms when faced with unfavorable environmental conditions (Lennon and Jones 2011). Our analysis also implies that the estimate of respiration rates under sufficient substrate by the physiological state index model is deficient. Pertaining to the switch function model, we argue that either the switch function ($\theta$) or the substrate (or combined with other impact factors) saturation level may be used to modify the microbial uptake rate but the inclusion of both is not only unnecessary but also inappropriate. Based on the generally accepted assumptions summarized from existing dormancy models, we postulate a synthetic microbial physiology component to account for dormancy. Both the steady state active fraction ($r^{ss}$) and substrate saturation level ($\phi^{ss}$) can be expressed as functions of two physiological indices: $\alpha$ and $\beta$. The index $\alpha = m_R / (\mu_G + m_R)$ is composed of $\mu_G$ and $m_R$ denoting the maximum specific growth and maintenance rates, respectively, for active microbes. The index $\beta$ represents the ratio of dormant to active maintenance rates. The value of $r^{ss}$ is no less than $\phi^{ss}$, and is equal only under the condition of $\beta \to 0$, where they are both identical to $\alpha$. The upper bound of $r^{ss}$ is ca. 0.8 at $\alpha \to 0.5$ and $\beta \to 1$. The maximum $r^{ss}$ is at the level of 0.5 if $\beta$ ($\leq 0.01$) following the estimation of Anderson & Domsch (1985a). It is evident that $r^{ss}$ could be attenuated further by other limiting factors. The application of the MEND microbial physiology model to an experimental dataset with intermittent substrate supply shows satisfactory model performance (the determination coefficients are 0.98 and 0.78 for microbial biomass and substrate, respectively). The case study on the SIGR dataset indicate that the exponentially-increasing respiration rates can only be used to determine $\mu_G$ and $B_{a0}$ (initial active biomass), while the major parameters ($B_0$, $r_0$, $\mu_G$, $\alpha$, $K_s$) can be effectively determined using both exponentially-increasing and non-exponentially-increasing respiration rates.

In conclusion, the microbial physiology model presented here can be incorporated into existing ecosystem models to account for dormancy in microbially-mediated processes. Traditional measures of microbial biomass include the entire microbial population, even though dormancy is an important evolutionary strategy for preservation of microbial genetics and function until conditions for growth and replication improve. Parameterizing microbial decomposition models assuming the entire population is active could therefore lead to significant errors. The approach described here provides a tractable and testable method to include dormancy as a response to external forcing.

**ACKNOWLEDGMENTS**

This research was funded by the Laboratory Directed Research and Development (LDRD) Program of the Oak Ridge National Laboratory (ORNL) and by the U.S. Department of Energy Biological and Environmental Research (BER) program. ORNL is managed by UT-Battelle, LLC, for the U.S. Department of Energy under contract DE-AC05-00OR22725. The authors thank Dr. Sindhu Jagadamma for her helpful comments.

# Appendix S1: A summary of two-microbial-pool models

1. Transformation between active and dormant states

| Reference | Model description |
|---|---|
| (Ayati 2012) | $B^N_{a \to d} = \gamma_{a \to d} \cdot B_a - \gamma_{d \to a} \cdot B_d$ <br> $\gamma_{a \to d} = \varepsilon_a / (S + K_a) + v_a$ <br> $\gamma_{d \to a} = \begin{cases} v_d, & \text{if } S > K_s \\ 0, & \text{otherwise} \end{cases}$ |
| (Bär et al. 2002) | $B^N_{a \to d} = v \cdot \{[1 - J(W - W_{ad})] \cdot B_a - J(W - W_{ad}) \cdot B_d\}$ <br> $J(W - W_{ad}) = 0.5 \cdot \tanh[(W - W_{ad})/s]$ |
| (Gignoux et al. 2001) | $B^N_{a \to d} = \begin{cases} (1-\delta) \cdot B_a - S^o/m, & \text{if } S^o < S^R \\ -(1-\delta) \cdot B_d, & \text{if } S^o \geq S^R \end{cases}$ <br> $S^R = m(1-\delta)(B_a + B_d)$ |
| (Hunt 1977) | $B^N_{a \to d} = \begin{cases} -\gamma_{d \to a} \cdot B_d, & \text{if } m \cdot B_a - S^o < 0 \\ 0 & \text{if } 0 \leq m \cdot B_a - S^o \leq \rho \cdot B_a \\ B_a - S^o/(\rho + m), & \text{if } m \cdot B_a - S^o > \rho \cdot B_a \end{cases}$ |
| (Jones and Lennon 2010) | $B^N_{a \to d} = (1 - R) \cdot B_a - R \cdot B_d$ <br> $R = R_{\max} \cdot \exp\{-(1 - W_f) \cdot [(E^*_L - E_L)/tol]^2 - W_f \cdot [(E^*_R - E_R)]^2\}$ |
| (Konopka 1999) | $B^N_{a \to d} = (1 - \mu/\mu_{\max}) \cdot \gamma_{a \to d} \cdot B_a - (\mu/\mu_{\max}) \cdot \gamma_{d \to a} \cdot B_d$ |
| (Locey 2010) | $B^N_{a \to d} = \gamma_{a \to d} \cdot B_a - (1 - p) \cdot (B_d/B) \cdot B_d$ |

**Variables and Parameters**:

$B_a$ : active biomass (mg C, mg C cm$^{-3}$, or mg C g$^{-1}$ soil, hereinafter referred to as mg C g$^{-1}$);

$B_d$ : dormant biomass (mg C g$^{-1}$);

$B^N_{a \to d}$ : net transformation of $B_a$ to $B_d$ (mg C h$^{-1}$, mg C cm$^{-3}$ h$^{-1}$, or mg C g$^{-1}$ h$^{-1}$, hereinafter referred to as mg C g$^{-1}$ h$^{-1}$);

$E_L$ and $E_L^*$: local environmental cue and its optimum;

$E_R$ and $E_R^*$: regional environmental cue and its optimum;

$J(W - W_{ad})$: probability for dormant bacteria to become active;

$K_a$ ($<K_s$): half-saturation constant for the conversion of active to dormant state (mg C g$^{-1}$);

$K_s$: half-saturation constant (mg C g$^{-1}$);

*m*: maintenance respiration rate (d$^{-1}$ or h$^{-1}$, hereinafter referred to as h$^{-1}$);

*p*: (1 − *p*) represents the probability of reactivation of dormant microbes;

*R* and *R*$_{max}$: resuscitation rate and maximum resuscitation rate (h$^{-1}$);

*s*: denote the steepness of the transformation function *J*(*W* − *W*$_{ad}$) in Bär *et al.* (2002);

*S*: substrate concentration (mg C g$^{-1}$);

*S$^o$*: potential offer of microbial substrate (mg C g$^{-1}$ h$^{-1}$);

*S$^R$*: maintenance respiration requirement (mg C g$^{-1}$ h$^{-1}$);

*tol*: environmental tolerance;

*v*: transformation rate between active and dormant states (h$^{-1}$);

*v$_a$* and *v$_d$*: rate constants (h$^{-1}$);

*W*: a stress field, e.g., soil humidity close to the surface in Bär *et al.* (2002);

*W$_{ad}$*: a critical value of the stress field, below which active bacteria incline to become dormant and vice versa;

*W$_f$*: weighting of local vs. regional environmental cues;

*ε$_a$*: rate constant (mg C h$^{-1}$);

*γ$_{a \to d}$* and *γ$_{d \to a}$*: transformation rates of active to dormant state and dormant to active state (h$^{-1}$);

*μ* and *μ*$_{max}$: specific growth rate and maximum specific growth rate (h$^{-1}$);

*ρ*: threshold rate that is set to 0.01 d$^{-1}$.

*δ*: microbial mortality rate (h$^{-1}$);

2. Switch function model

The switch function (Stolpovsky *et al.* 2011), i.e., determining the fraction (*θ*) of active biomass that uptakes substrate (e.g., DOC), follows a smoothed step function adapted from Fermi-Dirac statistics:

$$dB_a/dt = Y_{eff} \cdot \theta \cdot \mu(C_s, TEA) \cdot B_a - m_a \cdot B_a - (1-\theta) \cdot v_{deact} \cdot B_a + Y_{react} \cdot I_d \cdot \theta \cdot v_{react} \cdot B_d$$

$$dB_d/dt = -m_d \cdot B_d + (1-\theta) \cdot v_{deact} \cdot B_a - I_d \cdot \theta \cdot v_{react} \cdot B_d$$

$$dI_d/dt = I_d \cdot [k_{incr} \cdot \theta \cdot (1-\theta_I) - k_{decr} \cdot (1-\theta)]$$

$$\theta = \frac{1}{\exp\left(\frac{-G+G_0}{st \cdot G_0}\right)+1}$$

$$G = \Delta G \cdot \frac{1}{B_a} \cdot \frac{dS}{dt}\bigg|_{r=1, Y_G=0} = -\Delta G \cdot \mu(S, TEA)$$

Where $B_a$ and $B_d$ are the active and dormant biomass, respectively; $v_{react}$ and $v_{deact}$ are the specific rates for reactivation and deactivation, respectively; $Y_{react}$ (0.1–1) denotes the reactivation yield; $Y_{eff}$ is the effective growth yield; $m_a$ and $m_d$ represent the maintenance coefficient for $B_a$ and $B_d$, respectively; $I_d$ (0–1) represents the 'depth' of dormancy; $k_{incr}$ and $k_{decr}$ (0.05–0.5 h$^{-1}$) are the first-order rate constants describing the increase and decrease of $I_d$ under favorable ($\theta \rightarrow 1$) or unfavorable ($\theta \rightarrow 0$) conditions, respectively; $\theta_I$ is selected to ensure $I_d \leq 1$; $G$ is the maximum rate of Gibbs energy release per unit biomass; and $G_0$ (0.1–25 kJ mol$^{-1}$ biomass h$^{-1}$) represents a corresponding minimum threshold value; $\Delta G$ is the Gibbs energy change of the oxidation of substrate (e.g., oxidation of DOC into $CO_2$); $st$ is dimensionless and denotes the steepness of the step function; $\mu(S, TEA)$ is the growth rate as a function of concentrations of substrate (S) and terminal electron acceptor (TEA).

Stolpovsky *et al*. (2011) arbitrarily assigned a value of 0.1 to *st* that leads to a narrow but finite "switching zone". The value of $\Delta G$ is controlled by the concentration of substrate and products as well as the terminal electron acceptor (TEA).

## APPENDIX S2: Mathematical Derivations

1. Derivation of Eq. 4 based on Panikov (1995, 1996):
   From Eq. 1, it follows that

$$\frac{dB_a}{dt} = \frac{d(rB)}{dt} = r\frac{dB}{dt} + B\frac{dr}{dt} = r\frac{dB}{dt} + B \cdot \frac{1}{B}\frac{dB}{dt}(\phi - r) = \phi\frac{dB}{dt} \tag{S2-1}$$

2. Derivation of Eq. 12e: change of rate of active fraction ($r$)

$$\begin{aligned}\frac{dr}{dt} &= \frac{1}{B}\left(\frac{dB_a}{dt} - r\frac{dB}{dt}\right) \\ &= \frac{1}{B}\{(\phi/\alpha - 1) \cdot m_R \cdot B_a - (1-\phi) \cdot m_R \cdot B_a + \phi \cdot m_R \cdot B_d - r[(\phi/\alpha - 1) \cdot m_R \cdot B_a - (\beta \cdot m_R) \cdot B_d]\} \\ &= m_R\{(\phi/\alpha - 1) \cdot r - (1-\phi) \cdot r + \phi \cdot (1-r) - r[(\phi/\alpha - 1) \cdot r - \beta \cdot (1-r)]\} \\ &= m_R \cdot [(\phi - r) + (\phi/\alpha + \beta - 1) \cdot r \cdot (1-r)]\end{aligned} \tag{S2-2}$$

3. Steady state solution to the synthetic microbial physiology model (Eq. 12)

$$S^{ss} = K_S \frac{\alpha - \beta + 3\alpha\beta + \sqrt{C^2 + 8\alpha\beta}}{2(1-\alpha)(1+\beta)} \tag{S2-3a}$$

$$B^{ss} = \frac{Y_G I_S}{m_R} \cdot \frac{\alpha(-1+\beta)^2 + 3\beta + \beta^2 - (1-\beta)\sqrt{C^2 + 8\alpha\beta}}{4\beta^2} \tag{S2-3b}$$

$$B_a^{ss} = \frac{Y_G I_S}{m_R} \cdot \frac{C + \sqrt{C^2 + 8\alpha\beta}}{4\beta} \tag{S2-3c}$$

where $C = \alpha(\beta - 1) + \beta$

$$r^{ss} = \frac{B_a^{ss}}{B^{ss}} \tag{S2-3d}$$

$$\phi^{ss} = \frac{S^{ss}}{S^{ss} + K_S} \tag{S2-3e}$$

(1) $\beta = 0$

$$S^{ss} = K_s \cdot \alpha/(1-\alpha) \tag{S2-4a}$$

$$B^{ss} = I_s \cdot Y_G/(\alpha \cdot m_R) \tag{S2-4b}$$

$$B_a^{ss} = I_s \cdot Y_G / m_R \tag{S2-4c}$$

$$r^{ss} = \phi^{ss} = \alpha \tag{S2-4d}$$

(2) $\beta = 1$

$$S^{ss} = K_s \frac{4\alpha - 1 + \sqrt{1+8\alpha}}{4(1-\alpha)} \tag{S2-5a}$$

$$B^{ss} = I_s \cdot Y_G / m_R \tag{S2-5b}$$

$$B_a^{ss} = \frac{I_s \cdot Y_G}{m_R} \cdot \frac{1 + \sqrt{1+8\alpha}}{4} \tag{S2-5c}$$

$$r^{ss} = \frac{1 + \sqrt{1+8\alpha}}{4} \tag{S2-5d}$$

$$\phi^{ss} = \frac{4\alpha - 1 + \sqrt{1+8\alpha}}{3 + \sqrt{1+8\alpha}} \tag{S2-5e}$$

4. Derivation of Eq. 14.

   Simplification of Eq. 12(a–e) under conditions of (i) excess of substrate ($\phi \to 1$), (ii) $I_s = 0$, and (iii) the maintenance respiration of dormant microbes [$(\beta \cdot m_R) \cdot B_d$] may be negligible compared to the growth and maintenance respiration of active microbes.

$$dS/dt = -\frac{1}{Y_G} \cdot (\mu_G + m_R) \cdot (r \cdot B) \tag{S2-6a}$$

$$dB/dt = \mu_G \cdot (r \cdot B) \tag{S2-6b}$$

$$dr/dt = \mu_G \cdot r \cdot (1-r) + m_R \cdot (1-r) \tag{S2-6c}$$

Eqs. S2-6a–6c are integrated for initial conditions, i.e., $S = S_0$, $B = B_0$ and $r = r_0$ at $t = 0$:

$$S(t) = S_0 - \frac{B(t) - B_0}{Y_G (1-\alpha)} \tag{S2-7a}$$

$$B(t) = B_0 r_0 \cdot e^{\mu_G t} + B_0(1-r_0) \cdot [\alpha \cdot e^{\mu_G t} + (1-\alpha) \cdot e^{-m_R t}] \quad \text{(S2-7b)}$$

$$r(t) = \frac{[r_0 + \alpha(1-r_0)] \cdot e^{(\mu_G + m_R)t} - \alpha(1-r_0)}{[r_0 + \alpha(1-r_0)] \cdot e^{(\mu_G + m_R)t} + (1-\alpha)(1-r_0)} \quad \text{(S2-7c)}$$

The $CO_2$ production rate during the exponential growth stage, $v(t)$, is derived as an explicit function of $t$:

$$v(t) = \frac{dCO_2}{dt} = \frac{1-Y_G}{Y_G}(\mu_G + m_R) \cdot (r \cdot B)$$
$$= \frac{B_0(1-Y_G)}{Y_G}\{[(\mu_G + m_R) \cdot r_0 + m_R \cdot (1-r_0)] \cdot e^{\mu_G t} - [m_R \cdot (1-r_0)] \cdot e^{-m_R t}\} \quad \text{(S2-7d)}$$

If $m_R \ll \mu_G$ (i.e., $\alpha \to 0$), Eqs. S2-7a–7d are further simplified to

$$S(t) = S_0 - \frac{B(t) - B_0}{Y_G} \quad \text{(S2-8a)}$$

$$B(t) = B_0 r_0 \cdot e^{\mu_G t} + B_0(1-r_0) \quad \text{(S2-8b), similar to Eq. 11 in (Panikov and Sizova 1996)}$$

$$r(t) = \frac{r_0}{r_0 + (1-r_0) \cdot e^{-\mu_G t}} \quad \text{(S2-8c), similar to Eq. 10 in (Panikov and Sizova 1996)}$$

$$v(t) = \frac{B_0 r_0 (1-Y_G)}{Y_G} \mu_G \cdot e^{\mu_G t} \quad \text{(S2-8d), different from Eq. 13 in (Panikov and Sizova 1996) but}$$

similar to Eq. 7 in (Colores *et al.* 1996)